\documentclass[10pt,sigconf,letterpaper,nonacm,pbalance]{acmart}

\setcopyright{none}
\renewcommand\footnotetextcopyrightpermission[1]{}
\usepackage{booktabs}
\usepackage{array}
\usepackage{CJKutf8}
\usepackage{xspace}
\usepackage{enumitem}
\usepackage{graphicx}
\usepackage{booktabs,tabularx,array}
\usepackage{tikz}
\usetikzlibrary{arrows.meta,positioning,calc,fit,backgrounds,decorations.pathreplacing}

\colorlet{adred}{red!70!black}

\newif\ifshowdraft
\showdrafttrue
\ifshowdraft
  \newcommand{\pending}[1]{\textcolor{orange}{#1}}
  \newcommand{\todo}[1]{\textcolor{red}{\textsf{[TODO: #1]}}}
\else
  \newcommand{\pending}[1]{#1}
  \newcommand{\todo}[1]{}
\fi

\newcommand{\PkgFetched}{122{,}656}       
\newcommand{\PkgMissing}{52{,}051}        
\newcommand{\PkgLikely}{52}  
\newcommand{\PkgLikelyShare}{0.04\%}  
\newcommand{\PkgConfigurable}{18}  
\newcommand{\PkgLikelyOperatorAff}{48}  
\newcommand{\PkgRecoded}{76}
\newcommand{\PkgRecodeAgree}{68}
\newcommand{\PkgRecodeKappa}{0.76}         
\newcommand{\PkgPlaceOperator}{49}         
\newcommand{\PkgPlaceOperatorCode}{27}     
\newcommand{\PkgPlaceOperatorBackend}{22}  
\newcommand{\PkgPlaceLulu}{three}          
\newcommand{\PkgPlaceCodeEvidence}{30}     
\newcommand{\PkgPlaceCodeOperator}{29}     

\begin{document}

\title{Ads for Agents: Measuring Ad-Serving MCP Servers}
\author{Shahrooz Pouryousef}
\affiliation{%
  \institution{Chalmers University of Technology and University of Gothenburg}
  \city{Gothenburg}
  \country{Sweden}}

\author{Sohini Chakraborty}
\affiliation{%
  \institution{Stockholm University}
  \city{Stockholm}
  \country{Sweden}}

\author{Ali Montazer}
\affiliation{%
  \institution{Google}
  \country{USA}}

\author{Muoi Tran}
\affiliation{%
  \institution{Chalmers University of Technology and University of Gothenburg}
  \city{Gothenburg}
  \country{Sweden}}

\begin{abstract}
AI agents call external tools through protocols such as the Model Context Protocol (MCP).
Some MCP servers return paid advertisements (ads) into the model's context.
The model then decides whether these ads and their disclosures reach the user.
We trace these ads from their sources to the agent's tool results across public MCP servers and packages.
We find the servers through their public traces in ad networks, package registries, certificate logs, hostnames, MCP directories, and GitHub configurations.
We then query the ad-serving servers with read-only requests and inspect their ad links.
Ads appear on only {0.57\%} of 101{,}363 open MCP endpoints and \PkgLikelyShare{} of \PkgFetched{} local MCP packages.
Although {173} ad networks target agents, operators insert the ads themselves on about half of the 226 ad-returning servers.
Nearly all ad-serving packages likewise carry the operator's own affiliate identifier.
While {ads adapt to the topic and intent of the query}, their disclosure stays fixed within each tool.
Only 9.3\% of the ads carry an explicit sponsored label.
Most ad links also hide their destinations.
Since operators insert most ads themselves and label them inconsistently, we propose a per-item ad annotation in MCP results.
Clients can show this annotation to the user independently of the model's answer.
\end{abstract}



\maketitle


\section{Introduction}
\label{sec:introduction}

AI agents increasingly search, shop, and compare services for users by calling external tools.
The Model Context Protocol (MCP) standardizes these calls~\cite{mcpspec,anthropic2024mcp}.
In MCP, a server describes its tools and their input schemas.
The agent's model then chooses a tool and fills in its arguments.
The agent calls the tool and adds the server's response to the model's context.
Some MCP server operators monetize these responses with paid advertisements from services such as Lulu Ads~\cite{lulu2026mcp}.
For example, a shopping server can list a paid product marked \texttt{[Sponsored]} first.
We call such a paid item an \emph{ad}.

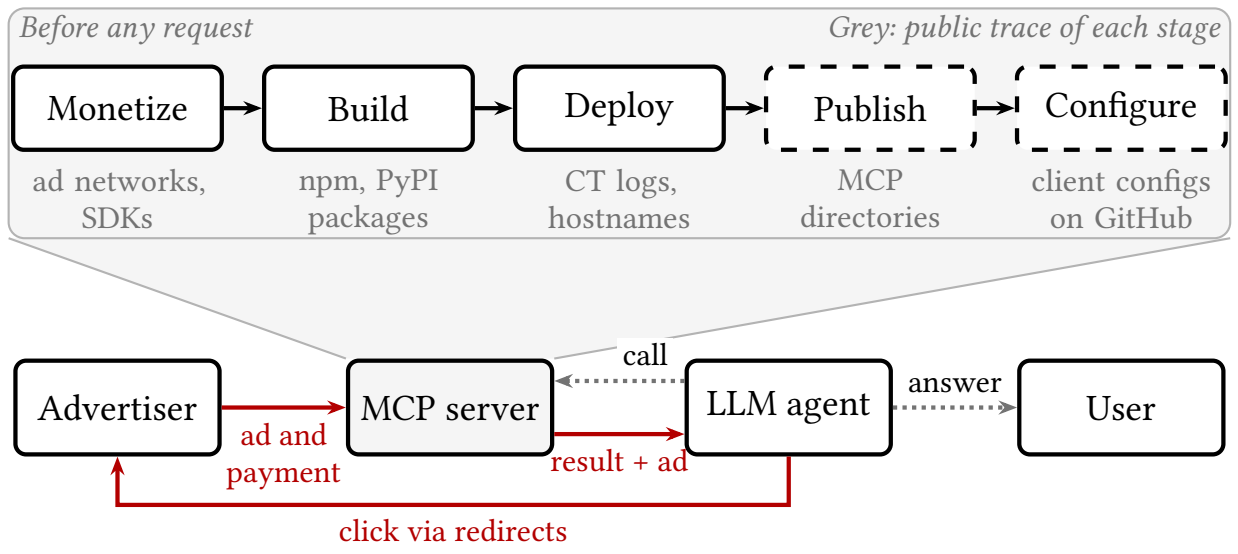
\begin{figure}[t]
\centering
\begin{tikzpicture}[
  font=\scriptsize,
  >={Stealth[length=3.5pt]},
  stage/.style={draw, thick, rounded corners=2pt, minimum width=1.38cm, minimum height=0.55cm, align=center, inner sep=1pt, fill=white},
  adstage/.style={stage, draw=adred, fill=adred!8},
  optstage/.style={stage, dashed},
  src/.style={font=\tiny, text=black!55, align=center, text width=1.45cm, inner sep=1pt},
  actor/.style={draw, thick, rounded corners=2pt, minimum height=0.62cm, minimum width=1.35cm, align=center, inner sep=2pt, fill=white},
  adactor/.style={actor, draw=adred, fill=adred!8},
  flow/.style={->, thick},
  adflow/.style={->, thick, draw=adred},
  offflow/.style={->, thick, densely dotted, draw=black!55},
  lbl/.style={font=\tiny, inner sep=1pt},
  rowlbl/.style={font=\tiny\itshape, text=black!55, inner sep=0pt},
]
\node[stage]    (mon) at (0,0)    {Monetize};
\node[stage]    (bld) at (1.66,0)  {Build};
\node[stage]    (dep) at (3.32,0)  {Deploy};
\node[optstage] (pub) at (4.98,0)  {Publish};
\node[optstage] (cfg) at (6.64,0)  {Configure};
\foreach \a/\b in {mon/bld, bld/dep, dep/pub, pub/cfg}
  \draw[flow] (\a) -- (\b);
\node[src, below=2pt of mon] {ad networks,\\ SDKs};
\node[src, below=2pt of bld] {npm, PyPI\\ packages};
\node[src, below=2pt of dep] (sdep) {CT logs,\\ hostnames};
\node[src, below=2pt of pub] {MCP\\ directories};
\node[src, below=2pt of cfg] {client configs\\ on GitHub};
\node[rowlbl, anchor=west] at (-0.66,0.52) {Before any request};
\node[rowlbl, anchor=east] at (7.30,0.52) {Grey: public trace of each stage};

\node[actor]   (adv)    at (0,-1.98)    {Advertiser};
\node[actor, fill=black!4] (server) at (2.2,-1.98) {MCP server};
\node[actor]   (agent)  at (4.44,-1.98) {LLM agent};
\node[actor]   (user)   at (6.64,-1.98)  {User};

\begin{scope}[on background layer]
  \coordinate (zTL) at (-0.72,0.66);  \coordinate (zTR) at (7.36,0.66);
  \coordinate (zBL) at (-0.72,-0.86); \coordinate (zBR) at (7.36,-0.86);
  \fill[black!4] (zBL) -- (server.north west) -- (server.north east) -- (zBR) -- cycle;
  \draw[black!30] (zBL) -- (server.north west);
  \draw[black!30] (zBR) -- (server.north east);
  \draw[black!30, fill=black!4, rounded corners=3pt] (zBL) rectangle (zTR);
\end{scope}

\draw[adflow]  (adv.east) -- node[lbl, below=1.5pt, text=adred, align=center]{ad and\\payment} (server.west);
\draw[offflow] ([yshift=5pt]agent.west) -- node[lbl, above=1.5pt, pos=0.3, fill=white]{call} ([yshift=5pt]server.east);
\draw[adflow]  ([yshift=-5pt]server.east) -- node[lbl, below=1.5pt, text=adred]{result + ad} ([yshift=-5pt]agent.west);
\draw[offflow] (agent.east) -- node[lbl, above=1.5pt]{answer} (user.west);
\draw[adflow]  (agent.south) -- ++(0,-0.32) -| (adv.south);
\node[lbl, below=1.5pt, text=adred] at ($(agent.south)!0.5!(adv.south) + (0,-0.32)$) {click via redirects};
\end{tikzpicture}
\caption{The path of an ad through MCP.
For each request (bottom), the ad travels to the agent, and a click returns through redirects.
The zoom shows the server's stages before any request (dashed stages are optional).
Red: the ad's path; dotted: out of scope.}
\Description{The bottom row shows one request from left to right: the advertiser passes an ad and a payment to the MCP server, the server returns a result with the ad to the LLM agent, which called the server, and the agent answers the user. A click on the ad leads from the agent through redirects back to the advertiser. A zoom above the MCP server shows the stages that the server passes through before any request: Monetize, Build, Deploy, Publish, and Configure, with the public trace of each stage below it in grey: ad networks and SDKs, npm and PyPI packages, certificate transparency logs and hostnames, MCP directories, and client configuration files on GitHub. Publish and Configure are dashed as optional. The call and the answer are dotted as out of scope.}
\label{fig:adpath}
\end{figure}

The operator prepares the path of an ad in stages before any request (Figure~\ref{fig:adpath}).
It first monetizes the server through its own placements, an ad network, or an affiliate program.
It then builds the server as a package and may deploy it as a remote server under a hostname with a TLS certificate.
It may also list the server in a directory.
Users then add the server to their MCP client's configuration.
At run time, the server returns ads to the agent in its tool results.
When the agent or the user follows an ad link, the click passes through redirects to the advertiser.
We call a server \emph{ad-serving} when it places ads in its responses.
Evidence of these ads can appear in the server's metadata, its source code, or its responses.

An ad in a tool result reaches the user only through the model.
Disclosure rules, however, assume the user sees each ad next to its label.
In the US, the Federal Trade Commission (FTC) requires a clear and conspicuous disclosure~\cite{ftc2015native}.
In the EU, Article~26 of the Digital Services Act (DSA) requires each ad to show that it is an ad, on whose behalf it appears, and why the user sees it~\cite{eu2022dsa}.
In MCP, this label is only text in the model's context.
The model therefore decides whether the ad and its label reach the user.
The agent also follows ad links without the browser defenses against redirect tracking~\cite{koop2020redirect}.

Whether an ad carries a label at all depends on who places it.
On the web, an ad network or a search engine often places and labels the ad uniformly.
Content creators who insert affiliate links, in contrast, often omit the disclosure~\cite{mathur2018endorsements}. In MCP, the operator, {an agent ad network, or an upstream service} can place the ad.
If a few ad networks place most ads, a disclosure standard must reach only these networks.
If operators place their own ads, it must reach every server.

Prior work tests ads for agents mostly in controlled experiments.
These experiments show that sponsored tags, list positions, and promotional wording shift which products and tools agents choose~\cite{allouah2025buying,wadi2026whom,wang2026advertising}.
Models instructed to favor a sponsor can also conceal the sponsorship from users~\cite{wu2026ads}.
These studies, however, test only researcher-designed ads.
The only audit of deployed chatbot ads that we know of finds that ChatGPT separates them from the generated text and shows them below the model's response~\cite{lurie2026chatgpt}.
Measurements of deployed MCP servers cover only their ecosystem and security~\cite{guo2025mcp,padilla2026exposed}.
As a result, it remains unknown which deployed MCP servers serve ads, who places them, and what reaches the agent when a server returns an ad.

We therefore follow the ad along its path from the operator's server to the user's agent:
\begin{itemize}[leftmargin=*,nosep]
\item \textbf{RQ1 (prevalence):} How prevalent is advertising among public MCP servers?
\item \textbf{RQ2 (supply):} Who places the ads on these servers?
\item \textbf{RQ3 (delivery):} What reaches the agent when a server returns an ad, and where does its link lead?
\end{itemize}

We tackle each research question as follows.
For RQ1, we build a census from the public traces of the setup stages in ad SDKs, package registries, certificate logs, MCP directories, and client configurations (Figure~\ref{fig:adpath}).
We then call the read-only tools of each open remote server with generated queries and statically read the code of each package.
Ad indicators in the metadata, responses, and code flag candidate servers and packages (Appx.~\ref{app:indicators}).
Independent language-model coders then confirm their ads with a common codebook.
For RQ2, we attribute each ad from its traces to the operator, {an ad network, or an upstream service}.
For RQ3, we test whether servers select ads for the query and repeat queries on the ad-serving tools with varied topic and intent.
For each ad, we check the five disclosure elements of the DSA and FTC guidance.
Without opening any ad link, we classify each link by the visibility of its destination.


We find that few public MCP servers serve ads and that their operators place these ads themselves.
These ads reach the agent with incomplete disclosure and behind links with hidden destinations.
Of {101{,}363} open remote endpoints, {579} ({0.57\%}) show evidence of ads.
Sponsored or affiliate content appears in the metadata of 495 of these servers and in our query results for 226.
Among packages, only \PkgLikely{} of \PkgFetched{} (\PkgLikelyShare{}) serve ads.
Although we identify {173} ad networks for agents, only {10} servers reference one of them in their metadata through an SDK package or an ad hostname.
Operators instead place the ads themselves on {117} of these {226} servers.
Likewise, \PkgLikelyOperatorAff{} of the \PkgLikely{} ad-serving packages carry the operator's own affiliate identifier.

The ads also depend strongly on the topic and intent of the user's query.
A repeated query returns nearly identical ads.
A query with another topic or intent, in contrast, returns substantially different ads (effects of 0.42 and 0.40, respectively).
Each tool's disclosure, however, stays the same {across all query intents}.
Of the 6{,}499 ads, 42.2\% show none of the five disclosure elements.
Only 9.3\% carry an explicit sponsored label.
Moreover, 64.9\% of the ad links conceal their destination, mostly behind opaque redirects on the operator's own host (41.0\%).
Because operators place most ads themselves, a disclosure standard must reach every server.
The only interface common to all servers is MCP itself.
We therefore propose to annotate each ad in MCP results with its label, advertiser, selection reason, and link destination (Section~\ref{sec:discussion}).
MCP clients can then show this annotation to the user independently of the model's answer.

\section{Data Collection}
\label{sec:data}

\begin{figure}[t]
\centering
\includegraphics[width=\columnwidth]{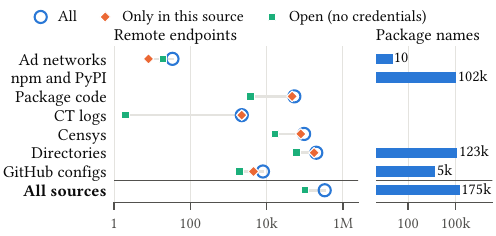}
\caption{Remote endpoints (left) and package names (right) per source (log scale).}
\label{fig:sources}
\end{figure}

\paragraph{Census.}
In November 2026, we collected the public traces of each setup stage of an ad-serving server (Figure~\ref{fig:sources}).
For the {\em monetize} stage, we searched the web, package registries, and GitHub for ad networks and affiliate services with placements in MCP tool results or LLM answers.
We stopped when two search rounds added no candidate.
We confirmed 173 of the 277 candidates against their documentation (Appx.~\ref{app:adnetworks}) and took their MCP server URLs, SDK names, and ad hostnames.
For the {\em build} stage, we took the npm and PyPI packages with ``mcp'' in their name and the MCP server URLs in their code.
For the {\em deploy} stage, we tried the \texttt{/mcp} and \texttt{/sse} paths on the hostnames with ``mcp'' in certificate transparency logs~\cite{rfc6962} and Censys.
For the {\em publish} stage, we crawled 41 of the 155 candidate directories and registries (Appx.~\ref{app:directories}).
For the {\em configure} stage, we extracted the servers from \nolinkurl{.cursor/mcp.json} and other client configuration files on GitHub.
We merged the records of each server by its endpoint URLs or its package name and linked endpoints to packages.
A server therefore runs remotely, ships as a local package, or both.
The census holds {332{,}130} remote-only servers, {151{,}497} local-only packages, and {21{,}129} servers with both.
Together, the remote servers expose {332{,}130} endpoints.

\paragraph{Probes.}
To read the metadata of the remote servers, we sent an unauthenticated MCP \texttt{initialize} handshake to 332{,}130 endpoints on November~12, 2026.
Of these endpoints, 101{,}363 (30.5\%) completed the handshake and listed their tools, prompts, and resources.
The 98{,}542 servers with a tool (Figure~\ref{fig:funnel}) expose 1{,}179{,}788 server--tool pairs (Table~\ref{tab:tool-funnel}).

To detect ads in tool responses, we called only safe tools.
These tools are read-only, accept synthetic arguments, and avoid private data and billable ad events (e.g., clicks or impressions).
A tool is read-only if it sets \texttt{readOnlyHint=true} or, without this annotation, has a read verb in its name.
Because MCP annotations are optional and untrusted~\cite{mcpspec}, we excluded tools with neither signal.

These criteria left 463{,}551 eligible tools (39.3\%) on 90{,}778 servers.
Of these tools, 303{,}665 qualify by annotation and the remaining 159{,}886 by name.
From the description and input schema of each eligible tool, a language model generated informational queries~\cite{broder2002taxonomy} and schema-valid arguments ({Appx.~\ref{app:probes}}).
We called 355{,}460 eligible tools (76.7\%) on 88{,}913 servers.
Of these tools, 214{,}186 (60.3\% of called and 46.2\% of eligible tools) on 79{,}892 servers returned at least one usable result.
The 108{,}091 uncalled eligible tools reflect coverage loss from unreachable servers and incomplete probing.
The prober abandoned a server after three consecutive errors.
Our logs omit the reason for each stop.

\paragraph{Code.}
On November~11, 2026, we fetched the latest release of each package in the census and verified its hash against the registry.
We excluded the \PkgMissing{} names without a release (30\%).
Most of these names are unpublished repositories listed in directories.
We read the remaining \PkgFetched{} releases statically in memory without installing or running them.

\section{How Many Servers Serve Ads?}
\label{sec:prevalence}
To answer RQ1, we search the census servers for advertisements in their metadata, tool responses, and package code.

\paragraph{Remote MCP servers.}
We identify ad-serving servers from their metadata and tool responses.
In the metadata, we search server instructions, tool descriptions, and schemas for advertising disclosures.
In the responses, we search for sponsored labels, affiliate identifiers, and known ad-network patterns (Appx.~\ref{app:indicators}).
Because these indicators only flag candidates, we code every server with a high-specificity indicator and random samples of those with weak or no indicators.
Independent LLM coders label each sampled server with one codebook (Appx.~\ref{app:codebook}).
The codebook separates advertising from self-promotion, ad management, and unrelated commercial content.
The coders label separately whether a server declares advertising and whether it returns advertisements.

\paragraph{MCP packages.}
We complement the probes with static analysis of npm and PyPI packages in our census.
We validate the package names, retrieve their published releases, and search the source code for advertising indicators.
In the code, we match dependencies, imports, launch commands, and referenced domains against known advertising SDKs, ad networks, and affiliate programs.
Because a local package exposes its tools only at run time, we also extract its tool definitions and search their descriptions and schemas.
We exclude matches confined to tests, examples, or bundled third-party code.
LLM coders then decide whether each flagged package inserts advertising into its tool results.
We count separately the packages that add advertisements only when the deployer supplies an advertising or affiliate identifier.

\begin{figure}[t]
\centering
\includegraphics[width=\columnwidth]{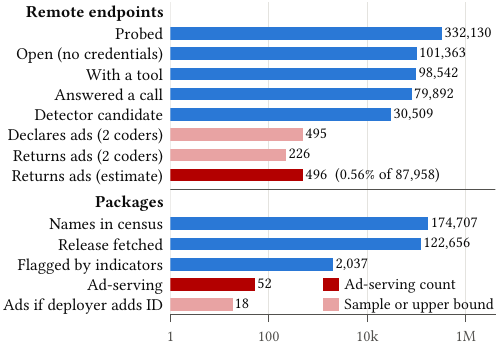}
\caption{Few public MCP servers serve ads. Remote endpoints and packages per step (log scale).}
\label{fig:funnel}
\end{figure}

\paragraph{Results.}
Few public MCP servers serve ads (Figure~\ref{fig:funnel}).
Our indicators flag 30,509 candidate servers, 20,737 by metadata and 12,488 by responses.
At least two LLM coders label 4,370 of the 14,558 sampled servers.
Both coders confirm ads in the metadata of 495 of these servers and in the tool results of 226.
Because 142 servers fall in both groups, 579 servers show evidence of ads.
Weighting the GPT-5-mini labels by inverse probability, we estimate that 496 of the 87,958 servers in our sampling frame (0.56\%; 95\% CI: 0.53--0.59\%) return ads.



\section{Who Places the Ads?}
\label{sec:supply}

To answer RQ2, we attribute each ad in tool responses and package code to the operator, an upstream API, or third-party content and classify operator ads as sponsored or affiliate.

\paragraph{Methodology.}
We attribute only ads with a concrete advertising trace.
The 226 ad-returning servers (Section~\ref{sec:prevalence}) return 6,644 candidate elements.
We keep candidates that match an affiliate, ad-network, or sponsorship indicator (Appx.~\ref{app:indicators}), a published affiliate URL pattern~\cite{mathur2018endorsements}, or a click path on the operator's host.
We discard candidates that match only a generic tracking parameter.
Deduplication by normalized destination leaves 5,163 ads on 226 servers.

Each inserter leaves a distinct trace.
We assign each ad to the first matching trace in a fixed order (Table~\ref{tab:rq2-def} in Appx.~\ref{app:attribution}).
Third-party content carries a provenance field whose source matches the link.
An upstream API leaves an ad-network MCP endpoint or SDK-backed link.
The operator leaves an own-host redirect, or an affiliate identifier or sponsorship marker backed by its monetization statement.
Because an affiliate identifier names only the commission earner, ads with only this identifier or no trace stay unknown.
Operator ads count as sponsored with a sponsorship marker and as affiliate with an affiliate identifier or own-host redirect.

\paragraph{Results.}
Operators insert 2,807 of the 5,163 ads (54.4\%) and are the best-supported inserter on 116 of 226 servers.
Upstream APIs and third-party content insert 205 ads (4.0\%) on 4 servers.
The remaining 2,151 ads (41.7\%) on 106 servers lack a known inserter.
Because operator insertion is easiest to establish, the counts may understate other inserters.

Both coding passes also attribute the ads in \PkgPlaceOperator{} of the \PkgLikely{} ad-serving packages to their operators.
In \PkgPlaceOperatorCode{} of these packages, the code itself adds the paid element (e.g., an Amazon tag).
The other \PkgPlaceOperatorBackend{} pass through affiliate links from the operator's backend.
Because this backend runs out of view, these attributions rest mostly on operator statements.
With code evidence alone, operators insert the ads in \PkgPlaceCodeOperator{} of \PkgPlaceCodeEvidence{} packages.
In the remaining \PkgPlaceLulu{} packages, all from ForgeMesh, the backend adds a sponsored card from the Lulu ad network.

\label{sec:mechanisms}
Most operator ads are unlabeled affiliate links.
Affiliate insertion accounts for {2,954} ads on {97} servers.
For example, \texttt{wem3.ai} wraps Awin links in a redirect on its own host.
The operator thus earns a commission on possibly organic results.
In the packages, advertisers pay mostly through their own referral programs (e.g., Amazon Associates) and less often through affiliate networks (e.g., Awin).
Sponsored insertion covers only {197} ads on {9} servers.
Because the operator adds these sponsored items itself, they can change which results the agent receives and how these results rank.

\section{What Reaches the Agent?}
\label{sec:delivery}

To answer RQ3, we measure how servers select ads for the user's query, how they disclose these ads to the agent, and where the ad links lead to.

\paragraph{Ad selection.}
\label{sec:rq3}
A server that selects ads for the query returns the same ads for a repeated query and different ads for a different query.
We test this property on {171} read-only ad-returning tools with a free-text query input (Section~\ref{sec:prevalence}).
Each tool receives two in-domain topics, each asked bare and with purchase intent (``buy \{t\} online'', {137} tools) or as a narrower request (``how does \{t\} work'', {34} tools).
We send the four queries in five rounds, 15 minutes apart, and open a fresh session per call.
Of the 3{,}540 calls, 95.1\% succeed.
To match an ad across calls, we decode its redirects and strip its click and session parameters.
We then compare the ad sets of two responses with the Jaccard similarity~$J$ (Appx.~\ref{app:selection}).

The returned ads depend on the query.
Repeated queries return almost the same ads ({$J=0.938$}).
Differences between queries therefore reflect the query itself.
The overlap falls significantly ($p=0.0001$), by 0.42 when the topic changes and by {0.40} when only the intent changes (Table~\ref{tab:rq3}).
A user who asks about the same subject with purchase intent thus receives different monetized links.
At the extreme, {11.4\%} of the responses contain no ad.

\begin{figure}[t]
\centering
\includegraphics[width=\columnwidth]{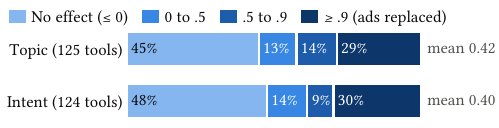}
\vspace{-0.7cm}
\caption{Share of tools by query effect: the drop in Jaccard similarity of their ads between repeated queries and queries with another topic or intent.}
\label{fig:rq3}
\end{figure}

This dependence holds for only part of the tools (Figure~\ref{fig:rq3}).
For 45\% of the tools, the topic has no effect on the ads.
In contrast, 29\% replace their ads almost entirely when the topic changes.
Monetization thus follows the tool in the first group and the query in the second.

\paragraph{Ad disclosure.}
\label{sec:ad-transparency}
The agent can recognize an ad only from its disclosure in the tool response.
Following the DSA and FTC guidance (Section~\ref{sec:introduction}), we check whether an ad carries a sponsored label, names the advertiser, explains its selection, states the commercial relationship, and separates itself from organic results~\cite{eu2022dsa,ftc2023endorsement,ftc2015native}.
We check these five elements for the 6{,}499 ads in the census responses of 226 servers.
We also classify the mechanism of each disclosure as a structured field, an inline text marker, server metadata, or none.

Only ads in structured fields disclose much of this information.
The five elements appear in 9.3\%, 30.0\%, 5.5\%, 28.7\%, and 6.5\% of the ads, in the order above.
An ad exposes 0.80 of the five elements on average and at most four.
Of all ads, 42.2\% expose none.
Ads in structured fields expose 3.00 elements on average, compared with 1.17 for ads with inline text markers (Appx.~\ref{app:disclosure-details}).
This gap partly reflects the format itself.
By construction, a structured field labels the ad and separates it from organic results.

The disclosure of each tool stays the same when the query changes.
Overall, explicit disclosure appears in {7.6\%} of the responses to general queries and in {16.4\%} of those to commercial queries.
Within each of the {100} tools with responses to both intents, however, the intent has no effect on disclosure ($p=1.0$).
The aggregate difference therefore reflects which tools return ads for each intent.
Servers thus adapt the ad to the query but keep its sparse disclosure fixed.

\paragraph{Ad links.}
\label{sec:destinations}
An agent that has not opened an ad link knows only what the link itself shows.
We therefore classify the 6{,}499 ad links by their URLs alone.
To separate first-party from third-party links, we compare the registrable domain of each link's initial host with the operator's domain.
To identify monetized links, we match each link against our inventory of 173 ad networks, published affiliate URL patterns~\cite{mathur2018endorsements}, and the HaGeZi referral list.
To recover hidden destinations, we recursively decode percent- and base64-encoded URLs in the path and in parameters such as \texttt{url=} (Appx.~\ref{app:link-details}).

Two thirds of the ad links hide their destination, mostly behind the operator's own redirect.
Only 35.1\% of the links show their destination, as a direct merchant URL (27.4\%) or encoded in the link (7.7\%).
First-party redirects on the operator's host, such as \texttt{operator.com/go/abc123}, form the largest opaque category (41.0\%).
Affiliate-network URLs without an encoded target make up the other 23.9\%.
An agent must open such a link to learn what it recommends.
In total, 46.1\% of the links are first-party.
The operator thus both attaches the monetization and controls the hop that hides the destination.
Ad networks with their own MCP server also appear as first-party.
For example, \texttt{mcp.feedico.io} serves 80 links on \texttt{partner.feedico.io}.
This host belongs to an inventoried ad network.
Even links with a visible destination often carry an affiliate parameter.
Of the 2{,}707 links (41.7\%) with an affiliate parameter, 1{,}353 are direct merchant URLs.
On these links, only the monetization remains implicit.


\section{Discussion}
\label{sec:discussion}

\begin{figure}[t]
\centering
\includegraphics[width=\columnwidth]{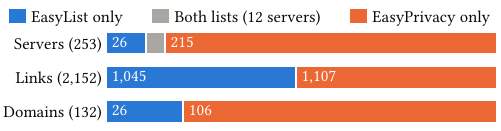}
\caption{Web blocklists match few MCP servers, mostly through tracking domains.
Each bar splits the matches of URL hosts in tool responses between \mbox{EasyList} (ads) and \mbox{EasyPrivacy} (tracking).
A matched host may serve tracking or ordinary content.}
\label{fig:blocklist}
\end{figure}

\paragraph{Disclosure in MCP\@.}
Web advertising blocklists recognize few of the ads in MCP tool responses.
We match the URL hosts in tool responses against the EasyList and EasyPrivacy versions of November 11, 2026.
Of the 79{,}892 servers with usable tool responses,
only 253 return at least one blocklisted URL (Figure~\ref{fig:blocklist}).
EasyList, the blocklist for advertising, matches only 292 of the 6{,}499 verified ad links (4.5\%).
This low match rate reflects where the ad links start.
Nearly half of these links start on the operator's own host (Section~\ref{sec:destinations}).
Many others start on merchant URLs with an affiliate parameter.
A host-based list blocks such links only by also blocking the ordinary content of the operator or the merchant.
The agent therefore depends on the server itself to mark which items are paid.

This mark belongs in MCP, the only interface common to all servers.
Because operators insert most ads themselves (Section~\ref{sec:supply}), a standard adopted only by ad networks would reach few servers.
We therefore propose a per-item ad annotation in the structured tool result.
Ads in structured fields already expose the most disclosure elements (Section~\ref{sec:ad-transparency}).
Placed in the structured result, the annotation separates the item from organic results and carries a sponsored flag, the advertiser, the selection reason, and the commercial relationship.
The selection reason accompanies each item and thus changes whenever the server adapts its ads to the query.
The annotation also states the destination of each ad link and thus reveals the 64.9\% of ad-link destinations hidden behind redirects.
A client can then display these fields next to the item regardless of the model's answer.
Clients and directories can also audit the operator's declarations against the affiliate parameters and operator redirects of the links in the same result (Section~\ref{sec:destinations}).

\paragraph{Limitations.}
Our measurements give a lower bound on advertising in MCP.
Our census covers only servers with a public trace (Section~\ref{sec:data}).
Servers behind credentials, {159{,}724} endpoints in our census, show their ads only to their users.
Our probes of public servers send fixed in-domain queries on few dates from one vantage point and therefore observe only the ads for these queries.

Our evidence also leaves payment and ad frequency unverified.
Within the observed ads, our attribution favors operators, the inserters with the clearest traces (Section~\ref{sec:supply}).
The 41.7\% of ads with an unknown inserter may therefore include ads from {ad networks and upstream services}.
The selection experiment uses synthetic queries without popularity weights and thus leaves ad frequency open.
Its two topics per tool, sent in one time window, establish only that the ads depend on the query.
The experiment replicates the carry-over control of~\cite{hannak2013} only in part and remains exposed to server state kept per IP address.
Similarly, static analysis shows which packages can insert ads but leaves open how many deployments activate this code (Section~\ref{sec:prevalence}).

\paragraph{Future work.}
Our measurement stops at the server's response.
Whether the user learns that an item is paid then depends on the model's answer.
The model receives the ad label only as text and can drop or rewrite it in its answer.
In controlled tests, models instructed to favor a sponsor conceal the sponsorship from the user~\cite{wu2026ads}.
A follow-up study can replay our responses through MCP clients and measure how often the label reaches the answer.

Data also flows from the agent to the server.
The agent fills tool arguments from the conversation and can pass the user's location or preferences to the server before any answer exists.
Prior work shows that agents place unnecessary personal data in tool arguments~\cite{li2026toolminimize} and that tool-call traces reveal user attributes~\cite{duan2026toolcalls}.
A follow-up study can test whether ad-serving servers request more context than comparable servers, forward it to third parties, or use it to select ads.

\section{Related Work}
\label{sec:related}

\paragraph{Ads and LLM agents.}
Commercial content in an LLM's context shifts which products and tools an agent chooses.
Crafted web pages and plugin documentation, for example, make LLM search engines promote an attacker's products~\cite{nestaas2025adversarial}.
In MCP, manipulative words in tool names and descriptions make agents prefer an attacker's server~\cite{wang2026mpma}.
The attacker then earns service fees or advertising income from the selected server.
List positions and promotional wording shift agent choices in the same way~\cite{allouah2025buying,wang2026advertising}.
A ``Sponsored'' label, by contrast, lowers the chance that an agent selects the labeled item~\cite{allouah2025buying,wang2026advertising}.
This penalty weakens when the system prompt names a booking platform as the agent's principal~\cite{wadi2026whom}.
Commercial influence also often escapes the user.
Models prompted to prioritize sponsors favor sponsored items and can conceal the sponsorship~\cite{wu2026ads}.
Users also often fail to recognize ads in chatbot answers, even when the answer carries a disclosure~\cite{tang2025adstalkback}.
In these studies, the researchers write the ads and the prompts.

Deployed and proposed ad systems differ in where they place the ad relative to the model.
ChatGPT shows its ads below the model's response, separate from the generated text~\cite{lurie2026chatgpt}.
A design proposal fetches ads through a tool call and tags them in the model's input~\cite{wu2025vigilant}.
Section~\ref{sec:discussion} proposes a comparable tag for each ad in MCP tool results.
Web agents meet ads in the pages they browse.
Attackers can buy such ads through ordinary ad delivery to inject malicious content into the agent's environment~\cite{wang2025adinject}.
This attack applies indirect prompt injection~\cite{greshake2023not} to ad delivery.
MCP servers place their ads in the tool results that the model reads.
We measure these deployed ads, their labels, and links (Sections~\ref{sec:prevalence}--\ref{sec:delivery}).

\paragraph{Measurements of MCP servers.}
Prior measurements locate public MCP servers in directories, package registries, and Internet scans to assess their quality and security.
Ecosystem studies collect 8{,}060 servers from six directories~\cite{guo2025mcp} and 67{,}057 servers from directories and the npm registry~\cite{li2026firstlook}.
Static analyses of open-source server code find vulnerabilities, tool poisoning, and privacy leaks~\cite{hasan2025mcp,yan2026leaks}.
Earlier studies of ChatGPT plugins and custom GPTs examine the tools of a single platform~\cite{iqbal2024llm,yan2024chatgpt,zhao2024gpts}.
One of these studies finds monetization terms on 48.7\% of the 396 reachable websites of GPT authors~\cite{zhao2024gpts}.
Closest to our census, studies of Internet-facing servers search scan engines and certificate transparency logs for remote endpoints~\cite{zhou2026authentication,padilla2026exposed}.
These studies confirm 7{,}973 and 640 live servers, respectively, through MCP protocol requests.
The second study also draws candidates from npm, PyPI, GitHub, and MCP directories~\cite{padilla2026exposed}.
Our census adds the traces of the monetize stage to these sources.
These traces include the MCP servers, SDKs, and ad hosts of 173 ad networks.
We confirm 101{,}363 endpoints through the MCP handshake (Section~\ref{sec:data}) and measure the ads in their tool results.

\paragraph{Web advertising measurement.}
Web ad measurement provides the methods behind our analysis of supply, disclosure, and ad selection.
Prior work extracts affiliate and merchant identifiers from URLs to attribute each click~\cite{chachra2015crookies}.
Recent work also follows the redirect chains of affiliate links and checks their disclosure against FTC rules~\cite{sun2026trust}.
Retargeted ads also reveal how ad exchanges share user data~\cite{bashir2016tracing}.
Disclosure research measures how people recognize native ads~\cite{wojdynski2016going} and how often content creators disclose their affiliate links~\cite{mathur2018endorsements}.
A recent framework operationalizes Articles 26 and 39 of the DSA to audit ad transparency~\cite{benzaamia2026dsa}.
Targeting studies supply our selection experiment's design.
One study compares treatment accounts with an identical control account and finds personalization in 11.7\% of Google results~\cite{hannak2013}.
AdFisher runs browser experiments with significance tests and finds fewer ads for high-paying jobs for female profiles~\cite{datta2015}.
We apply these methods to MCP ads (Sections~\ref{sec:supply} and~\ref{sec:delivery}).

Ad links, in turn, pass through redirect chains.
Studies of malicious ads and redirect tracking reconstruct the chains of ad delivery and clicks~\cite{zhou2012madtracer,koop2020redirect}.
Open redirects let attackers extend such chains to any destination~\cite{ndssopenredirect}.
On deal-sharing sites, 67.9\% of affiliate redirect chains add more tracking techniques~\cite{roux2026gotta}.
Redirectors that pass user identifiers also often escape tracker lists~\cite{randall2022uid}.
Request-based lists such as EasyList and EasyPrivacy also miss tracking data in link decoration~\cite{munir2024purl}.
These methods assume that a person views the ad in a browser.
In MCP, the model reads the ad and its label.
The agent follows its links without a browser.


\section{Conclusion}
\label{sec:conclusion}

We measure advertising in public MCP servers and packages.
Few of these servers serve ads.
Their operators insert most of the ads themselves, mainly as affiliate links and redirects. {They adapt these ads to the query's topic and intent} but keep their disclosure fixed and incomplete.
Most ad links also hide their destination.
Agents therefore receive ads without consistent information about their sponsorship or destination.

We expect ads for agents to become more common.
Ad networks already sell placements in tool results and model answers.
Our census captures this early market and offers a systematic baseline for its growth.
As the market grows, more ads reach agents before they reach users.
Ad transparency therefore has to travel with each result through the agent's protocols.
To this end, we propose a standardized, per-item ad annotation in MCP results for clients to display independently of the model's answer.
Sponsorship then becomes part of each result, like its content and source.
Such an annotation lets clients show users which results are paid and lets directories audit operators at scale.
Advertising for agents can then grow as transparently as advertising for people on the web.

\bibliographystyle{ACM-Reference-Format}
\bibliography{references}

\appendix

\interlinepenalty=10000
\raggedbottom

\section{Ethics}
\label{app:ethics}

Our measurement sends rate-limited requests only to public MCP servers and uses no accounts or credentials.
The census identifies our client by name and sends only the MCP handshake and the requests listing tools, prompts, and resources.
The probes send one in-domain query to each safe tool (Section~\ref{sec:data}) and repeated calls only to servers with ad signals (Section~\ref{sec:delivery}).
To avoid billing advertisers, we click no ad link.
Because crawled text can contain instructions for a language model, no model acts on it.

\section{Open Science}
\label{app:openscience}

We release the server census with each server's sources, the ad-serving labels, the codebooks, the full indicator lists with the ad-network SDKs (Appendix~\ref{app:indicators}), and the analysis scripts at \href{https://anonymous.4open.science/r/mcp-ad-measurement-7D02/}{\textcolor{ACMDarkBlue}{anonymous.\allowbreak 4open.\allowbreak science/\allowbreak r/\allowbreak mcp-ad-measurement-7D02}}.
We redact credentials, tokens, and affiliate click identifiers from the released URLs and tool responses.

\section{Generative AI}
\label{app:genai}

Language models write the probe arguments, confirm the ad networks, code the packages, and estimate the ads missed by our indicators.
For the probes, a language model writes each tool's arguments from its name, description, and input schema under one fixed prompt (Appendix~\ref{app:probes}).
Opus 5.5 validates them against the schema before each call.
For the ad networks, two coders, ChatGPT 5 and Opus 5.5, confirm each candidate against its documentation and packages (Section~\ref{sec:data}).
For the packages, agents apply our codebook in two independent passes (Section~\ref{sec:prevalence}).
Their prompt tells them to read each release only as text and to ignore its instructions.
The two passes agree on \PkgRecodeAgree{} of the \PkgRecoded{} packages with ads ($\kappa=\PkgRecodeKappa{}$).
We resolve the rest by reading the code.
For the missed ads, a language model checks each server's metadata and responses against our ad definition alone (Section~\ref{sec:prevalence}).

We draft scripts and edit prose with an AI coding assistant.
The authors review and take responsibility for its output.

\begin{table*}[t]
\centering\footnotesize
\setlength{\tabcolsep}{4pt}
\caption{Most of the 173 confirmed ad networks for agents serve ads only in chat apps and agent clients. Rows group the networks by how their ads reach the user. Networks in italics have shut down.}
\label{tab:adnetworks}
\begin{tabular}{@{}>{\raggedright\arraybackslash}p{0.125\textwidth}>{\raggedright\arraybackslash}p{0.86\textwidth}@{}}
\toprule
Own MCP server (27) &
Adgentek, AdoraAds, Advertising.chat, Agent Offer Network (AON), AgentClick.ai, AgentFuse, \textit{agentic-ads}, bestbid.lol, \textit{Boost Boss (Lumi)}, BuyWhere, \textit{Buzzer Ads}, Clickwise, Cuelinks, DefaultBench, Feedico, Godalo, IntentLink, Kone, Mirakl Ads (MCP-native ad serving), MoltAd, Nexbid, picoads, PPToGo, Rako (Agent Attribution Protocol), Rollhub / Agent Casino affiliate, RootVine, Syndicate Links \\
\addlinespace[3pt]
SDK or API for agents and MCP servers (39) &
AdCrier, AdIntelli, AdLib, AdMesh, \textit{ADS4GPTs}, Adsshare (maket mcp-bidder), Adstract AI, AdZen, AffiliateSwarm, AgentAds, \textit{AgentLoop}, AgentVine, AttentionMarket, Aura / Affinity AI Bid API, \textit{Bido}, Channel3, ChatAds, \textit{cl1s.tech}, Clad AI, Dexter x402 sponsored-access, \textit{EarnLayer}, EthicalAds, Fill (fill.md), \textit{Flynnstone}, Imprezia, Lulu, \textit{Monetzly}, \textit{NativeAds}, NotPixel, OpenAds.AI, \textit{Operon}, \textit{Pyrimid}, Smalk AI Ads, sponsoredtokens, Taboola DeeperDive, VADS, Wildfire, Zeekend, \textit{ZeroClick} \\
\addlinespace[3pt]
Chat apps and agent clients only (107) &
ad402 (sponsormyterminal), adback, adline, Adly/Rowadly (myk794), adMarketplace AI Discover, adsline, AdsOverAI, adspay, ADtention, advibing, Adviverse, AIBC, aicash (Gravity reseller), Amplora, Aryel, Awaitful, Backfill, BackSpin, Bramble, \textit{Ceed Ads}, ChatWait, ClaudeCash, Clickback, \textit{CodeCash}, Codevertise, Conversaic, Dappier, deadtime, deals-dev, dilinx, Doceree Admanager (AI Ads on Chat Screen), Elo (formerly Growl), Firsthand, Garlic, getlull, Goodwait (ImpactWait), Google Business Agent for Leads, Gravity, gravy, \textit{idleads}, idleai, IdleDev, Idlen, idlepay, \textit{IdlePay (idlepay.dev)}, \textit{IdlePay (idlepay.xyz)}, iwantads, Jutera, Kickbacks, kickbacksai (npm), Koah Labs, Kontext, Kontezza, Kovio, \textit{Latent Protocol}, LoadEarn, \textit{Lullad}, Lumaline, MatrixAds, Meanwhile, Microsoft Advertising (Copilot ads), Minefy Seller, moula, Nexad, Obrigado, OpenAI ChatGPT Ads, \textit{opencode-ads (Asphalt9bla)}, OpenCrater, OUTRIK AI Buffer Ads, \textit{PayloadAds}, \textit{Perplexity Ads}, Platfory, \textit{Platita}, \textit{Prism Ad Network}, promptai.credit, PromptBid, \textit{ProRata Gist Ads}, Ramtro, \textit{Sidecar}, Simula, Spinr, Spinwage, \textit{Spinyield}, Sponsored Code, SponsoredAI, Sponsy, Teads, \textit{TermAds}, thespin.ad, Thrad, TICK, Tickover, tipbar, TokenWait, \textit{Tributly}, uWait, Velocity, VibeAd, Vibefuel, Waitji AI, WaitLine (getwaitline.com), WaitLine (waitline.dev), WaitMint, WaitPerk, WaitRoll, WaitSpin, wavebird \\
\bottomrule
\end{tabular}
\end{table*}

\section{Data Collection}
\label{app:data}

\paragraph{Ad networks.}
\label{app:adnetworks}
Table~\ref{tab:adnetworks} groups the 173 ad networks that we confirm (Section~\ref{sec:data}) by the route through which their ads reach the user.

\paragraph{Directories.}
\label{app:directories}

We searched for directories in the places where users and MCP clients look for servers.
These places are the web, GitHub, articles that compare directories, the aggregator list of the official MCP Registry, and the catalogs of MCP clients.
From each verified candidate, we also followed its badges and footer links to other directories.
When two consecutive search batches added no qualifying directory, we stopped with 155 candidates on November~6, 2026.

A candidate enters the census only when it adds servers that we can probe or fetch.
Such a directory lists at least 100 third-party MCP servers, is online without login, allows automated crawling, gives a remote URL or a package for each entry, and duplicates no other directory.
Of the 155 candidates, 41 meet these criteria (Table~\ref{tab:directories}).
Of the 114 others, 39 are enterprise registries, single-vendor catalogs, or plugin directories.
The rest are offline or need a login (24), duplicate another directory (23), list fewer than 100 servers (15), or block crawlers (13).
Crawler blocks exclude the two largest, PulseMCP (21{,}587 servers) and mcpservers.org (19{,}479).
We re-checked all decisions and changed four to apply the duplicate and plugin rules uniformly.

\begin{table}[t]
\centering\footnotesize
\caption{The census crawls 41 of the 155 candidate directories. These directories span official registries and client catalogs, commercial directories, and curated lists. GitHub lists appear under their owner's name.}
\label{tab:directories}
\begin{tabular}{@{}>{\raggedright\arraybackslash}p{\columnwidth}@{}}
\toprule
\textit{Official registry and client catalogs (8):} Official MCP Registry, GitHub MCP Registry, Docker MCP Catalog, Microsoft MCP Center, Claude Connectors, Cline Marketplace, ToolHive Registry, ModelScope MCP Square \\
\addlinespace[3pt]
\textit{Commercial directories (15):} Glama, Smithery, Stork, mcp.so, MCP.Directory, MCPBundles, LobeHub, ToolSDK, MCP Server Finder, MCP Market, Ever Works, Etropo, MCPize, AIBase, Shyft \\
\addlinespace[3pt]
\textit{Curated lists and registries (18):} AI Anytime, mcpHQ, MCPM, HeyClaude, aitmpl, Archestra, TensorBlock; GitHub lists of wong2, appcypher, YuzeHao2023, MobinX, VoltAgent, best-of-mcp-servers, Awesome-MCP-ZH, habitoai, rohitg00 (DevOps), zackchewa (China), jaw9c (remote) \\
\bottomrule
\end{tabular}
\end{table}

\paragraph{Tool probing.}
\label{app:probing}
\label{app:probes}
Table~\ref{tab:tool-funnel} traces the tools of the 98{,}542 open servers from eligibility to a usable result.
Of the 1{,}179{,}788 tools on these servers, 463{,}551 (39.3\%) meet our eligibility criteria (Section~\ref{sec:data}).
We attempted to call every eligible tool and reached 355{,}460 of them (76.7\%).
Of the called tools, 214{,}186 (60.3\%) returned at least one usable result.
The uncalled eligible tools therefore reflect coverage loss.
Some servers became unreachable at call time.
On servers that responded only partially or failed repeatedly, the prober stopped before it reached all eligible tools.

\begin{table}[!tb]
\centering\footnotesize
\setlength{\tabcolsep}{4pt}
\caption{Of the 463{,}551 eligible tools, 214{,}186 (46.2\%) return a usable result to our probes. Each tool is a (server, tool name) pair. The excluded rows and the eligible row sum to all tools.}
\label{tab:tool-funnel}
\begin{tabular}{@{}lrrr@{}}
\toprule
\textbf{Stage} & \textbf{Tools} & \textbf{\%} & \textbf{Servers} \\
\midrule
\multicolumn{4}{@{}l}{\emph{Endpoint discovery}} \\
Endpoints probed                &               &                 & 332{,}130 \\
Open endpoints                  &               &                 & 101{,}363 \\
With $\geq$1 tool               &               &                 & 98{,}542 \\
\midrule
\multicolumn{4}{@{}l}{\emph{Tool eligibility}} \\
All tools                       & 1{,}179{,}788 & 100.0\%         & 98{,}542 \\
\quad Excluded: not read-only   & 377{,}870     & 32.0\%          & \\
\quad Excluded: read-only unknown & 186{,}958   & 15.8\%          & \\
\quad Excluded: arguments not fillable & 93{,}370 & 7.9\%         & \\
\quad Excluded: private or ad-event name & 58{,}039 & 4.9\%       & \\
\textbf{Eligible}               & \textbf{463{,}551} & \textbf{39.3\%} & \textbf{90{,}778} \\
\quad by \texttt{readOnlyHint=true} & 303{,}665  & 25.7\%          & \\
\quad by read verb in the name  & 159{,}886     & 13.6\%          & \\
\midrule
\multicolumn{4}{@{}l}{\emph{Tool calls}} \\
\textbf{Called}                 & \textbf{355{,}460} & \textbf{76.7\%}$^{a}$ & \textbf{88{,}913} \\
\textbf{With $\geq$1 usable result} & \textbf{214{,}186} & \textbf{60.3\%}$^{b}$ & \textbf{79{,}892} \\
\bottomrule
\multicolumn{4}{@{}l}{\scriptsize $^{a}$\,Of eligible tools. \quad $^{b}$\,Of called tools (46.2\% of eligible tools).}
\end{tabular}
\end{table}

\section{Ad Detection}
\label{app:detection}

\paragraph{Indicators.}
\label{app:indicators}
A server can carry an ad in its metadata, its tool responses, or its code and links.
Tables~\ref{tab:ind-meta}--\ref{tab:ind-net} list the seed indicators for each place.
Each indicator only flags a candidate.
{An author then codes} each candidate with the codebooks.

The indicators combine the labels of observed ads with those prescribed by regulators and ad networks.
In the Source column, P marks the ad terms in our pilot crawl, R the disclosure terms of the FTC and the EU~\cite{ftc2015native,eu2022dsa}, T their translations, N the documentation of the confirmed ad networks, and A our added disclosure phrasings.
Some terms also match content without ads.
For example, ``ad'' matches Active Directory, ``commission'' the European Commission, and ``sponsor'' clinical trials.
The coding removes these matches.


\paragraph{Sampling.}
\label{app:sampling}
The indicators also set the sampling stratum of each remote server.
We code all high-specificity servers and random samples of the other strata (Section~\ref{sec:prevalence}).
Tables~\ref{tab:ind-meta}--\ref{tab:ind-net} mark each indicator group as high-specificity~(H) or weak-signal~(W).
We also count the affiliate URL patterns of~\cite{mathur2018endorsements} as high-specificity and the EasyList and EasyPrivacy domains as weak-signal.
Each server enters the strongest stratum among its matches.
Servers with a successful tool result and no match form the no-signal stratum.

\paragraph{Server codebook.}
\label{app:codebook}
An item counts as an ad only when it shows a paid relationship.
Such an item is marked sponsored or promoted, contains a verifiable affiliate link (e.g., Amazon \texttt{tag=}, Awin, or eBay \texttt{campid}), or is identified as affiliate or referral content in the response or server documentation.

A commercial result without such evidence counts as organic content.
Coders apply this definition to one advertising signature at a time.
A signature is the link or label mechanism of one tool on one server.
To find the signatures, the coders inspect the server metadata and the recorded tool responses.
Because opening a link can register an affiliate click, the coders leave links unopened.
For each server, the coders label separately whether it declares ads and whether it returns them.


\paragraph{Package codebook.}
The package codebook applies the same test to code.
It asks whether the runtime code of a package adds paid content to its MCP tool results.
Paid content is an item or a link paid for by a third party.
Examples are an ad from an ad network, a sponsored item, and an affiliate or referral link.
Organic results, such as products from an upstream API, count as paid content only when the code shows an affiliate parameter or host, a sponsored flag, or an operator statement about commissions.

The coding unit is a package's latest release on November~11, 2026.
For each release, the coder reads the scanner matches, the package files, the registry description, and the README as text and installs or runs nothing.
Evidence of paid content must come from runtime code.
Runtime code is the package's code outside tests, examples, documentation, and vendored directories such as \texttt{node\_modules}.

Each package receives the first applicable status of Table~\ref{tab:codebook-pkg} and, if it has paid content, a group.
The group records how the paid content reaches the agent.
Because Section~\ref{sec:prevalence} counts only paid content in a package's own tool results, only ad-serving packages in the core group count.
Section~\ref{sec:prevalence} reports the configurable packages of this group separately.


\section{Ad Attribution}
\label{app:attribution}

\paragraph{Servers.}
For each ad on a remote server, coders identify the inserter, the party that added the advertising element.
Each ad receives the first category of Table~\ref{tab:rq2-def} whose evidence it matches.
An operator that wraps an upstream result in its own affiliate link, redirect, or sponsored flag counts as the inserter.
The evidence ranks from source code to response traces, documentation, and finally operator-hosted redirects alone.
Domain names, promotional keywords, and affiliate-tag text establish the inserter only together with other evidence.
The attribution names the inserter of an ad and leaves its payee open.

\paragraph{Packages.}
For each ad-serving package, the coder records the placing party, the mechanism, the evidence level, and the paying party.
The placing party, which decides which ad appears, is {the operator or its own backend, an ad network, a third-party service, or the deployer.}
The mechanism is a call to an ad network, an affiliate identifier or redirect, or a sponsored listing.
A package can use several.
The evidence level is code when the runtime code or bundled data shows the paid element, an identifier when only the documentation shows it, and a statement when only the operator says that a link earns a commission.
The paying party is the advertiser's own referral program, an affiliate network, an ad network, or an advertiser buying a featured slot directly.
The confidence of each code is high when the code shows the mechanism directly and medium when one step, such as a remote backend, is out of view.

\section{Ad Delivery}
\label{app:delivery}

\paragraph{Ad selection.}
\label{app:selection}

The query experiment of Section~\ref{sec:rq3} keeps the queries independent of the ads.
Each of its {177} tools passed a per-tool semantic review.
The topics of each tool come from a per-domain pool.
A hash of the tool identity selects two of these topics.
Therefore, the topics are independent of the tool's earlier ads.
We froze the query set (SHA-256 \texttt{e61aaa21}) before the first call.
Each round permutes the query order from a recorded seed.
Each server receives one call at a time with all other arguments fixed.

The ad sets of two responses determine each similarity value.
We compare the ad sets of two responses with the Jaccard similarity~$J$.
Because $J$ is undefined for two empty sets, we exclude these pairs.
Empty ad sets make up {11.4\%} of the responses.
The effect of topic or intent is the within-query similarity minus the similarity between queries that differ in this factor (Table~\ref{tab:rq3}).
A paired permutation test over tools tests each effect against zero.
Cluster bootstraps over tools give the 95\% confidence interval of each effect.
We adapt seed-based query generation from~\cite{schultheiss2023}, the repeated measurement and the control-versus-control noise floor from~\cite{hannak2013}, and randomization with a distribution-free permutation test from~\cite{datta2015}.

\paragraph{Ad disclosure.}
\label{app:disclosure-details}
The five disclosure elements of Section~\ref{sec:ad-transparency} are a sponsored label, the advertiser's identity, a selection explanation, the commercial relationship, and separation from organic results.
No ad exposes all five.
Only four ads expose four of them.
The number of exposed elements depends on the disclosure mechanism.
Ads in structured fields expose 3.00 elements on average, compared with 1.17 for inline text markers, 0.44 for no explicit mechanism, and 0.26 for server metadata.
Inline markers often state the commercial relationship (75.8\%) but rarely separate the ad from organic results.
Each tool also keeps one mechanism.
All {125} examined tools keep a constant disclosure flag in their responses.


\paragraph{Ad links.}
\label{app:link-details}
The URL of an ad link is our only evidence about it.
Because opening a link can register a click, we leave links unopened and observe no landing page.
A link counts as first-party when the registrable domain of its initial host matches the operator's domain.
We match the hosts against our 173 confirmed ad networks, published affiliate patterns~\cite{mathur2018endorsements}, the HaGeZi referral list, EasyList, EasyPrivacy, and tracker databases.
We also recursively decode the destinations embedded in parameters such as \texttt{url=} and \texttt{redirect=}.
Ad and tracker blocklists match a small share of the 6{,}499 links.
EasyList matches 292 links (4.5\%) and EasyPrivacy 845 (13.0\%).
Tracker databases attribute 1{,}817 links (28.0\%) to known organizations.
The visible destinations include \texttt{expedia.com} (83 links), \texttt{amazon.com} (34), and \texttt{ebay.de} (31).


\begin{table*}[p]
\centering\scriptsize
\setlength{\tabcolsep}{4pt}
\caption{Keyword indicators for the metadata: server instructions, the names and descriptions of tools, prompts, and resources, and the field names in input and output schemas and \texttt{\_meta}. A slash separates alternatives after one prefix. Stratum: high-specificity~(H), weak-signal~(W), or recorded only~(--).}
\label{tab:ind-meta}
\begin{tabular}{@{}>{\raggedright\arraybackslash}p{0.13\textwidth}>{\raggedright\arraybackslash}p{0.73\textwidth}cc@{}}
\toprule
Group & Terms & Source & Stratum \\
\midrule
Sponsored placement & sponsored result/listing/product/placement/slot/block/content/recommendation/link; promoted result/listing/product/placement/post/item; paid placement/listing/featured/promotion/ranking; featured placement; paid promotion; [sponsored] & P & H \\
Ad business & ad slot/unit/network/inventory/placement/block; native ad; agent-ad; advertiser near agent, LLM, AI, or assistant; CPM; CPC; CPA; revenue share; earn per recommendation/click/impression/call; monetize near MCP, agent, tool, or server; billing token & P & H \\
Affiliate & affiliate link/URL/code/ID/program/commission/tag; referral fee/code/link/commission/bonus/reward; commission on/from/per/when; we earn/receive/get a commission/fee/referral; kickback; pays [name] a referral & P & H \\
Field names & \texttt{sponsored}, \texttt{is\_sponsored}, \texttt{promoted}, \texttt{paid\_placement}, \texttt{premium\_placement}, \texttt{ad}, \texttt{ads}, \texttt{advert*}, \texttt{ad\_id}, \texttt{ad\_request\_id}, \texttt{ad\_label}, \texttt{ad\_disclosure}, \texttt{ad\_slot}, \texttt{ad\_unit}, \texttt{ad\_block}, \texttt{ad\_creative}, \texttt{ad\_impression*}, \texttt{ad\_click*}, \texttt{ad\_placement}, \texttt{ad\_campaign\_id}, \texttt{billing\_token}, \texttt{affiliate*}, \texttt{commission*}, \texttt{referral\_code/link/url/fee/id}, \texttt{sponsor}, \texttt{sponsor\_label/disclosure/name/id}, \texttt{featured}, \texttt{is\_featured}, \texttt{disclosure*} & P & H \\
Neutrality claims (recorded, not ads) & no, never, not influenced by, without, or free of, followed by paid placement, sponsored, affiliate bias, pay-to-rank, or advertising; never paid; no ads & P & -- \\
Disclosure & advertisement; advertising; ads; paid; paid partnership; material connection; we may earn; earns a commission; \#ad & R & W \\
Paid ranking and sponsor & paid inclusion; paid ranking; higher ranking; payment for ranking; on behalf of; paid for by; sponsored by & R & H \\
Ad-network vocabulary & sponsored; sponsored by; sponsored via; suggestion from; placement; offer; impression; bid; auction; campaign; publisher; monetize; revenue share & N & W \\
Common phrasings & partner offer; promoted by; in partnership with; brought to you by; presented by & A & H \\
Translations & German: Anzeige, Werbung, gesponsert, Partnerlink, Provision; French: publicit\'e, sponsoris\'e, lien affili\'e, partenariat r\'emun\'er\'e; Spanish: publicidad, patrocinado, anuncio, enlace de afiliado; Portuguese: publicidade, patrocinado, an\'uncio, link de afiliado, comiss\~ao; Dutch: advertentie, gesponsord; Chinese: \begin{CJK}{UTF8}{gbsn}广告, 赞助, 推广, 佣金, 返利\end{CJK}; Japanese: \begin{CJK}{UTF8}{min}広告\end{CJK}, PR, \begin{CJK}{UTF8}{min}スポンサー, アフィリエイト\end{CJK}; Korean: \begin{CJK}{UTF8}{mj}광고, 협찬, 제휴, 수수료\end{CJK} & T & W \\
\bottomrule
\end{tabular}
\end{table*}

\begin{table*}[p]
\centering\scriptsize
\setlength{\tabcolsep}{4pt}
\caption{Label indicators for the tool responses. Parentheses name the network that documents each format. H/W: high-specificity in responses and weak-signal in the metadata; translations are weak-signal in both.}
\label{tab:ind-resp}
\begin{tabular}{@{}>{\raggedright\arraybackslash}p{0.13\textwidth}>{\raggedright\arraybackslash}p{0.73\textwidth}cc@{}}
\toprule
Group & Labels & Source & Stratum \\
\midrule
Item flags & \texttt{is\_sponsored: true}; \texttt{"sponsored": true} or an object; \texttt{is\_promoted: true}; \texttt{"promoted": true}; \texttt{"is\_ad": true}; \texttt{"ad": true}; \texttt{promotion\_tier}; \texttt{sponsored\_rank}; \texttt{disclosure\_label: "ad"}; \texttt{"featured": true}; \texttt{tier}, \texttt{placement}, \texttt{listing\_type}, or \texttt{badge} set to featured, promoted, sponsored, or paid & P & H \\
Text markers & [Sponsored]; (Sponsored); sponsored via; a heading that starts with Sponsored; \texttt{<!-{}- agent-ad}; paid placement; Anzeige; Werbung & P & H \\
Ad tools & a tool whose name contains advert, sponsor, promot, ad, ads, adunit, or placement and that returns a non-empty result & P & W \\
Response disclosures & affiliate link; Amazon Associate; earn a commission; may earn; commission from/on; sponsored & P & H \\
Negative markers & \texttt{is\_sponsored: false}; \texttt{"sponsored": false} or empty; \texttt{"placements": []}; no paid placement; not sponsored & P & -- \\
Network formats & \texttt{placement.sponsored: true} (Operon); \texttt{disclosure: "Sponsored"} with \texttt{disclosure\_label: "Ad"} (AdoraAds); \texttt{"sponsored": \{"label": "Sponsored"\}} (Lulu); \texttt{slot.disclosure} (Zeekend); \texttt{disclosure} with \texttt{label}, \texttt{explanation}, and \texttt{sponsor\_name} (AttentionMarket); \texttt{disclosure: "sponsored"} (agentic-ads); \texttt{rel="sponsored"} (PromptBid, ChatAds); Ad (AdMesh, Imprezia); Sponsored by [advertiser] (Thrad, Imprezia, mcpadserver); Sponsored via picoads (picoads); Suggestion from Kontext (Kontext); Sponsored Blips $\cdot$ [sponsor] (OpenCrater) & N & H \\
Added flags & \texttt{ad\_id}; \texttt{ad\_type}; \texttt{advertiser}; \texttt{sponsored\_by}; \texttt{sponsor\_name}; \texttt{campaign\_id}; \texttt{bid}; \texttt{boosted}; \texttt{paid}; \texttt{partner: true}; \texttt{label} or \texttt{tag} set to Ad, Sponsored, Promoted, or Partner & A & {H/W} \\
Added text markers & Ad $\cdot$; Advertisement; Promoted; Paid partnership; Featured partner; \#ad; the translations in Table~\ref{tab:ind-meta} & A, T & {H/W} \\
\bottomrule
\end{tabular}
\end{table*}

\begin{table*}[p]
\centering\scriptsize
\setlength{\tabcolsep}{4pt}
\caption{Ad-network and affiliate indicators for the code and the links in responses. Strata apply to the links.}
\label{tab:ind-net}
\begin{tabular}{@{}>{\raggedright\arraybackslash}p{0.13\textwidth}>{\raggedright\arraybackslash}p{0.73\textwidth}cc@{}}
\toprule
Group & Indicators & Source & Stratum \\
\midrule
Affiliate parameters & Amazon \texttt{tag=[id]-2x}; \texttt{aff}, \texttt{affiliate}, \texttt{affiliate\_id}, \texttt{affid}, \texttt{aff\_id}, \texttt{aff\_sub}, \texttt{subid}, \texttt{sub\_id}, \texttt{clickref}, \texttt{cjevent}, \texttt{irclickid}, \texttt{awinaffid}, \texttt{campid}, \texttt{ascsubtag}, \texttt{camref}, \texttt{ref\_id}, \texttt{refid}, \texttt{referral}, \texttt{referral\_code}, \texttt{ref\_code}, \texttt{partner\_id}, \texttt{affcid}, \texttt{aid}, \texttt{pid}, \texttt{mkcid}, \texttt{tracking\_id}, \texttt{associate}, \texttt{ranmid}, \texttt{ransiteid} & P & H \\
Paid campaign tags & \texttt{utm\_medium} set to cpc, ppc, paid, paidsearch, affiliate, display, sponsored, ad, or ads & P & H \\
Affiliate and ad hosts & amzn.to, evyy.net, sjv.io, pxf.io, ojrq.net, awin1.com, prf.hn, anrdoezrs.net, dpbolvw.net, kqzyfj.net, tkqlhce.com, jdoqocy.com, shareasale.com, linksynergy.com, skimresources.com, redirectingat.com, rstyle.me, s.click.aliexpress.com, tp.media, tp.st, travelpayouts.com, rover.ebay.com, partnerize.com, howl.me, geni.us, hop.clickbank.net, digistore24.com, viglink.com, sovrn.com, tradedoubler.com, webgains.com, impact.com, pjtra.com, pjatr.com, emjcd.com, commission-junction.com, flexoffers.com, admitad.com & P & H \\
Redirect and attribution (only with another indicator) & paths /go/, /out/, /aff/, /track/, /click, /clk, /redirect, /redir, /rs/, /visit/, /ref/, /r/; other \texttt{utm\_} parameters; \texttt{src}, \texttt{source}, \texttt{via}, \texttt{agent}, \texttt{channel}, \texttt{surface}, \texttt{s}; destinations nested in \texttt{u}, \texttt{url}, \texttt{to}, \texttt{d}, \texttt{dest}, \texttt{destination}, \texttt{ued}, \texttt{aff\_url}, \texttt{target}, \texttt{redirect}, \texttt{link}, \texttt{r}, \texttt{out}, \texttt{goto}, or in base64 path tokens & P & W \\
Ad-network SDKs (npm, PyPI) & 47 packages, e.g., @operon/sdk, @kontextso/sdk-js, @gravity-ai/api, admesh, ads4gpts-core, promptbid, chatads-sdk, @picoads/eliza-plugin; full list in the released data (Appendix~\ref{app:openscience}) & N & H \\
Ad-network hosts & 51 hostnames, e.g., api.operon.so, server.trygravity.ai, exchange.promptbid.ai, ssp.thrad.ai, api.useadmesh.com, api.dappier.com, picoads.xyz; full list in the released data (Appendix~\ref{app:openscience}) & N & H \\
\bottomrule
\end{tabular}
\end{table*}

\begin{table*}[p]
\centering\footnotesize
\caption{Codebook for package code. The coder assigns the first status that applies. Ad-serving and configurable packages also receive a group. The Packages column counts the flagged packages per code after both coding passes, as ad-serving / configurable for the groups. Only ad-serving packages in the core group count toward prevalence.}
\label{tab:codebook-pkg}
\begin{tabular}{@{}>{\raggedright\arraybackslash}p{0.17\textwidth}>{\raggedright\arraybackslash}p{0.66\textwidth}r@{}}
\toprule
Code & Use when & Packages \\
\midrule
\multicolumn{3}{@{}l}{\emph{Status}} \\
Ad-network own & The package belongs to an ad network or affiliate program, such as its publisher SDK or its own MCP server. Excluded as a server of the ad network. & {68} \\
Ad-serving & Runtime code adds paid content to the tool results by default, for example through a call to an ad network, a hard-coded or default affiliate tag, a sponsored flag, or an affiliate redirect. & {96} \\
Configurable & Runtime code adds paid content only when the deployer supplies an identifier. Reported separately. & {23} \\
Ad management & The tools manage ads, campaigns, or affiliate programs for an account holder, such as an advertiser. Excluded. & {427} \\
Promotional & The package promotes its author's own product or service, with no paid relationship. & {35} \\
Mention only & An indicator matches, but the package does not monetize its tool results, for example in documentation, a catalog of ad networks, or an API schema. & {626} \\
False positive & The match has nothing to do with ads, such as Active Directory or the sponsor of a clinical trial. & {762} \\
\addlinespace
\multicolumn{3}{@{}l}{\emph{Group (ad-serving and configurable only)}} \\
Core & The package, its bundled data, or its operator's backend adds the paid content to the package's own tool results. Counted. & \PkgLikely{} / \PkgConfigurable{} \\
Platform pass-through & A scraper or API wrapper returns the source platform's own labeled ads. & {20 / 2} \\
Incidental & Bundled third-party content, such as a copied list, carries another party's tag. & {9 / 0} \\
Transaction referral & A referral identifier travels only in a sign-up or payment request, which the agent does not read. & {9 / 3} \\
Host channel & Paid content reaches the agent's host through hooks or a status line, outside the tool results. & {2 / 0} \\
Not an MCP server & The package exposes no MCP tools, such as a library or an agent plugin. & {4 / 0} \\
\bottomrule
\end{tabular}
\end{table*}

\begin{table*}[p]
\centering\footnotesize
\caption{Inserter categories in test order. Each ad receives the first category whose evidence it matches. An \emph{operator statement} is the server's own text saying that it adds affiliate, referral, or sponsored content.}
\label{tab:rq2-def}
\begin{tabularx}{\textwidth}{@{}l X@{}}
\toprule
\textbf{Category} & \textbf{Accepted evidence} \\
\midrule
Third-party content
  & A provenance field (\texttt{source}, \texttt{provider}, \texttt{via}, \dots) attributes the element to a retrieved page on another domain.
    The element lies outside the operator's host.
    The operator's text is silent on adding such links. \\
\addlinespace
Upstream API
  & An API or feed supplies the element.
    The server is a published MCP endpoint of a confirmed ad network, or the link points to a network host and the server's package carries that network's SDK. \\
\addlinespace
Operator
  & The operator produces the element.
    The evidence is a redirect or click \emph{path} on the operator's own host, or an affiliate identifier, published affiliate pattern, or sponsored marker \emph{together with} an operator statement. \\
\addlinespace
Multiple parties
  & Two parties contribute the same element.
    This category requires the upstream response.
    Because we did not store upstream responses, no ad receives this category. \\
\addlinespace
Unknown
  & An ad with an affiliate identifier or sponsored marker and no further evidence of its inserter. \\
\midrule
\multicolumn{2}{@{}l}{\emph{Evidence strength}} \\
Strong   & Evidence independent of the operator's claims (own-host redirect, network endpoint, SDK, provenance field). \\
Moderate & A URL identifier or emitted label plus an operator statement. \\
Weak     & An identifier or label alone. \\
\bottomrule
\end{tabularx}
\end{table*}

\begin{table*}[p]
\centering\small
\caption{Effect of the query on the returned ads: Jaccard similarity~$J$ of repeated queries minus $J$ of queries with another topic or intent.
Brackets: 95\% cluster-bootstrap CI; all $p=0.0001$.}
\label{tab:rq3}
\begin{tabular}{@{}lcc@{}}
\toprule
Second intent & Purchase ({137} tools) & Narrower ({34} tools) \\
\midrule
$J$ of repeated queries & 0.928 & 0.989 \\
Topic effect  & \textbf{0.416} \scriptsize[.33,.50] & \textbf{0.450} \scriptsize[.24,.66] \\
Intent effect & \textbf{0.382} \scriptsize[.30,.47] & \textbf{0.510} \scriptsize[.30,.72] \\
\bottomrule
\end{tabular}
\end{table*}

\end{document}